# Hyperuniformity in amorphous speckle patterns

Diego Di Battista[1,2,3,*], Daniele Ancora[1,2], Giannis Zacharakis[1], Giancarlo Ruocco[4], and Marco Leonetti[5,6]

[1]*Institute of Electronic Structure and Laser, Foundation for Research and Technology-Hellas, N. Plastira 100, Vasilika Vouton, 70013, Heraklion, Crete, Greece.*

[2]*Materials Science and Technology Department, University of Crete, 71003, Heraklion, Greece*

[3]*Assing S.p.A., Monterotondo, Rome, Italy*

[4] *Center for Life Nano Science@Sapienza, Instituto Italiano di Tecnologia, Viale Regina Elena, 291 00161 Roma (RM) Italia.*

[5] *CNR NANOTEC-Institute of Nanotechnology c/o Campus Ecotekne, University of Salento, Via Monteroni, 73100 Lecce, Italy*

*\*dibattista.d@iesl.forth.gr*

**Abstract:** Hyperuniform structures possess the ability to confine and drive light, although their fabrication is extremely challenging. Here we demonstrate that speckle patters obtained by a superposition of randomly arranged sources of Bessel beams can be used to generate hyperunifrom scalar fields. By exploiting laser light tailored with a spatial filter, we experimentally produce (without requiring any computational power) a speckle pattern possessing maxima at locations corresponding to a hyperuniform distribution. By properly filtering out intensity fluctuation from the same speckle pattern, it is possible to retrieve an intensity profile satisfying the hyperuniformity requirements. Our findings are supported by extensive numerical simulations.

## 1. Introduction

Disorder [1] is generally considered detrimental for imaging or optical information transport, nevertheless, complex systems can disclose extraordinary physical properties, allowing the confinement of light [2], and enabling high resolution imaging techniques [3].

The task of driving light at the nanoscale is usually fulfilled by homogeneous waveguides or photonic crystals supporting a band gap [4]: completely ordered structures which allow to tailor the light flow. On the other hand, recent studies on complex materials and photonic glasses [5, 6] demonstrated that complete gaps can be found also in case of designed disorder [7, 8], which is today inspiring new platforms to overcome the present limitations encountered in photonic applications [9] and enabling extraordinary optical phenomena [10-13].

Light propagating in disordered media generates speckle patterns that are also complex light features, they are originated by random scattering through or by a rough dielectric slab. These light structures can be tailored to produce semi-organized light patterns [14-16] able to constitute random potential-energy landscapes [17, 18] in use for particles [19] or ultracold atoms [20] manipulation and with striking properties if adopted for structured illuminations [21, 22]. The Amorphous Speckle Patterns (ASP) are a kind of light structures with a short-range correlated light arrangement, generated by a superposition of Bessel modes [13] that lends the pattern the self-healing nature [22], an advantageous property for enhancing optical imaging in thick tissues [22, 23]. Besides that, ASP can be exploited to fabricate static structures with disordered refractive index microstructure, as required to observe Anderson localization [24-26] and are recurring features in non-linear phenomena like rogue waves [27] and random lasing [28]. For these reasons ASPs are attracting large interest in the field of Optics and Photonics.

On the other hand, Anderson localization is often associated with hyperuniform dielectric structures and their direct correlation is currently examined [29, 30]. Hyperuniform systems [31] are exotic states of matter fabricated exploiting designed disorder laying between a crystal and a liquid: they behave similarly to a perfect crystal since they suppress large scale density fluctuation, but they also are statistically isotropic with no Braggs peaks as in liquid and glasses [31-33]. They have been observed in different physical systems ranging from disordered ground state [34], to jammed particle packing [35], in processes of ultracold atoms [36] and in driven non-equilibrium systems [37]. A statistically homogeneous hyperuniform points configuration in $d$-dimensions is one in which the number variance $\sigma_N^2(R)$ of points within a spherical observation window of radius $R$ grows more slowly than $R^d$

$$\sigma_N^2(R) \sim R^{d-1} \qquad (1)$$

or, equivalently, it possesses a structure factor that satisfies the following condition as the wavenumber $k \equiv |\boldsymbol{k}|$ [33]:

$$\lim_{|\boldsymbol{k}| \to 0} S(\boldsymbol{k}) = 0. \qquad (2)$$

If there is no scattering in an exclusion region around the origin in $\boldsymbol{k}$-space, i.e.,

$$S(\boldsymbol{k}) = 0 \quad for\ 0 \leq |\boldsymbol{k}| \leq K \qquad (3)$$

a hyperuniform system becomes "stealthy" [34, 38, 39].

The hyperuniform structures are currently subject of study in the regulation of light confinement associated with their photonic band gaps [29, 30, 40] and are exploited for photonic applications at the micro-scale [40-44]. Nevertheless, the creation of hyperuniform heterogeneous material [38] requires very strict fabrication protocols resulting very challenging and costly, even though modern lithographic methods are able to produce super defined hyperuniform structures [45-47].

Hyperuniformity has been also generalized to heterogeneous media [33, 38] and random scalar fields [39]. In the latter case the hyperuniformity concept can be generalized by replacing the $S(\boldsymbol{k})$ with the spectral density $\tilde{\psi}(k)$ [39]. The hyperuniform condition can be simply extended to scalar fields that respect the following equation:
$$\lim_{|\boldsymbol{k}|\to 0} \tilde{\psi}(k) = 0. \tag{4}$$
Currently, the challenge is on the identification of analogue real system capable to replicate hyperuniform continuous field distribution theorized with the recent works [33, 39].

In this paper we analyze ASP and investigate on their connection with hyperuniform continuous fields supporting our study with rigorous numerical simulations.

We demonstrate that the hyperuniform distribution of the ASP intensity peaks satisfy the conditions in equations (1) and (4). To do so we examine normalized auto-correlation function, $\tilde{\chi}(r)$, and the spectral density $\tilde{\psi}(k)$ of tailored light fields distributions.

Moreover, we demonstrate that it is possible to obtain a continuous variable hyper uniformity (by applying an opportune intensity filtering in Fourier domain) from a standard laser speckle.

## 2. Looking for hyperuniform continuous fields

In order to identify an optical equivalent that best replicate a continuous hyperuniform field, we simulate tailored distribution of non-random light fields and we examine synthetic speckle patterns directly designed from a given hyperuniform distribution. This preliminary study serves to extrapolate the parameters that characterize a hyperuniform speckle pattern and we intend to use those as starting point to build a more realistic hyperuniform continuous field distribution in our experiments.

The easiest way to simulate a hyperuniform continuous field is by adopting Gaussian phasors [48] spatially placed in a [X,Y] grid according to the coordinates of a given hyperuniform matrix (the matrix has been provided by Authors of Ref. [34]). We analyse the speckle patterns generated from the superposition of 800 (number of points in the given hyperuniform matrix) Gauss modes ($f_{i=1,\dots,800}$) distributed in the grid according to a given $\chi = 0.4$ stealthy hyperuniform map. The dimensionless parameter $\chi$ represents the ratio between the number of constrained degrees of freedom and the total number of degrees of freedom that have been set to build the hyperuniform map [34]. In practice, we replace the points of a hyperuniform distribution with interfering Gaussian phasors/modes with the aim to form a continuous filed distribution. Considering that the average distance between neighbouring points in the hyperuniform map is $\tilde{d} = 368 \text{pixels}$ (calculated as the distance between the peaks in $\tilde{\chi}(r)$ from the hyperuniform map), we compared the speckle patterns resulting from superposition of modes with $\sigma/\tilde{d} = 0.054$ (Fig. 1a), 0.24 (Fig. 1b), 0.43 (Fig. 1c), 0.86 (Fig. 1d).

We first study the variation in $\tilde{\chi}(r)$ (line *II* of Fig. 1) and $\tilde{\psi}(k)$ (line *III* of Fig. 1) for the cases of $\sigma < \tilde{d}$ corresponding to Fig. 1a, b and c. Those patterns are secured (by the map) to maintain the hyperuniform geometry; they are three continuous fields exhibiting hyperuniform distributions as proved with the observation of their $\tilde{\psi}(k)$ reported in the line *III* of Fig. 1a, b and c. For the three cases, we have that $\tilde{\psi}(k) \to 0$ when $k \to 0$, meaning that the hyperuniform condition of Eq. (4) is respected.

When Gaussian sources are much smaller than their average distance, interference is negligible, nevertheless when they start consistently overlapping the interference between the fields becomes relevant and the hyperuniform structure is lost, this is the case of $\sigma \geq \tilde{d}$ in Fig. 1d. Here, the hyperuniform from the map information is completely lost and the final speckle assumes the typical random distribution as demonstrated with the analysis of $\tilde{\chi}(r)$ and $\tilde{\psi}(k)$ shown in lines *II* and *III* of Fig. 1d.

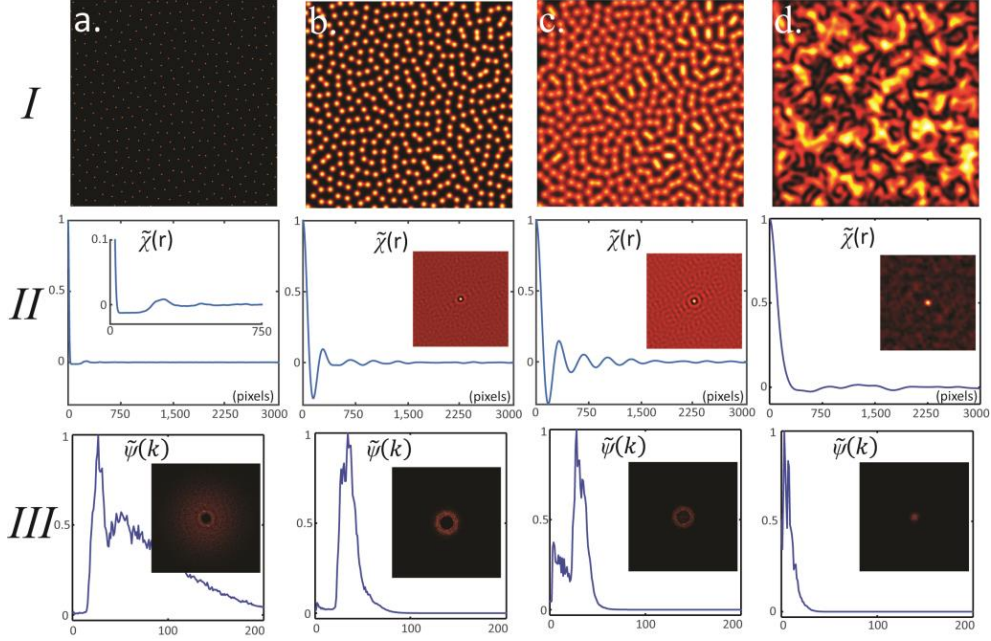

Fig. 1. Gauss modes distributed according to the $\chi = 0.4$ stealthy hyperuniform map. We compare the $\tilde{\chi}(r)$ and $\tilde{\psi}(k)$ from the patterns composed of modes with different width $\sigma$: in panel a. the $\sigma$ is 20 pixels, in b. is 90 pixels, in c. is 60 pixels and in d. is 320 pixels. The distance between the first two peaks in the $\tilde{\chi}(r)$ from c. provides the average distance between the modes position, $\tilde{d} = 368$ pixels. While in a., b. and c. $\sigma \ll \tilde{d}$ and the hyperuniform information from the map is maintained ($\tilde{\psi}(k) \to 0$ when $k \to 0$), in d. we have $\sigma \geq \tilde{d}$ the speckle pattern results completely disordered.

Therefore, the condition $\sigma \ll \tilde{d}$ must be respected to maintain hyperuniformity from points to continuous field distributions. Although this first intelligible analysis might result trivial, it discloses a very important piece of information: we have the evidence that as larger the $\sigma$ (in the regime $\sigma \ll \tilde{d}$) as the $\tilde{\chi}(r)$ tends to a Bessel function, meaning that hyperuniformity for continuous fields requires a Bessel shaped auto-correlation function of the speckle distribution as confirmed by the statistical analysis illustrated in lines *II* and *III* of Fig. 1c. This is a statement that finds also theoretical support in recent studies demonstrating that continuous hyperunformity is well described with Bessel-like auto-correlation function [39]. From this fundamental concept we base our strategy to build hyperuniform continuous fields in our simulations and experiments.

### 2.1 Superposition of Bessel modes

Standard speckle patterns result from interference of scrambled light fields, they represent a random distribution of maxima and minima in intensity. Their Fourier analysis is very well established [50]: correlation function with a sharp peak at short radius (around $r = 0$) and decaying spectral density for larger wavenumbers (see Methods). Differently, ASP exhibit Bessel-shaped auto-correlation function and a discontinuous spectral density: very bright around $k = 0$ and null at larger wavenumbers unless those residing in a very tight interval around $k = k^*$ with $k^* \gg 0$ [13].

In this section we report numerical and experimental results demonstrating that the statistic of a superposition of Bessel beams is independent on their particular arrangement differently from what has resulted with the previous case illustrated in Fig. 1.

In our experiments, in order to generate ASPs, we apply a spatial frequency selection to the light transmitted through a scattering barrier, a filtering operation performed by an annular aperture (see Methods). Exploiting the annular filter we generate a superposition of plane waves with fixed $k$ and random phases, which is known to give rise to an ASP.

On the other hand, in order to obtain a numerical comparison with our experiments, we simulate synthetic ASPs with superposition of Bessel modes/dephasors as described in equation 5 below:

$$J_{\alpha,i} = \left(\frac{z}{2}\right)^\alpha \sum_{k=0}^{\infty} \frac{\left(\frac{-z_i^2}{4}\right)^k}{k!\Gamma(\alpha+k+1)} \cdot e^{i\cdot\varphi_i} \quad (5)$$

With the integer order $\alpha = 0$, $\Gamma$ is the gamma function and $z_i = 2\pi * s * R(x - x_i, y - y_i)$. The magnitudes $s$, $R_i(x_i, y_i)$ and $\varphi_i$ represent respectively the size and the position of the mode in [X,Y]=5000x5000 pixels grid, the position of the i-th mode and the phase associated. In this case, the resulting pattern is given by the following formula:

$$C_{simul.}^{Bessel} = \left(\sum_{i=1}^{N} J_{\alpha,i}\right)^2 \quad (6)$$

A combination of dephasors defined in such a way, describes the experimental condition (see Fig. 6b in Methods), in which the interference pattern on the camera plane results from a superposition of Bessel beams.

In Fig. 2 we compare ASPs generated exploiting interfering Bessel beams. In Fig. 2a (simulation 1) we are summing randomly arranged Bessel sources [$x_i$ and $y_i$ in $R(x - x_i, y - y_i)$ are random], in 2b (simulation 2) sources are located on the grid according to a given hyperuniform map [33] (same method described for Gauss modes in Fig. 7 of Methods) and in 2c we report the ASP from our experiment (where phasors location is completely random). First we note that ASP generated from the hyperuniform map (Fig. 2b) does not differ qualitatively from the ASP produced with Bessel modes distributed in random positions (Fig. 2a) or from the experimental ASP of Fig. 2c. This is confirmed when we study their distributions in terms of $\tilde{\chi}(r)$ and $\tilde{\psi}(k)$; they exhibit same pair statistics. For all the patterns, the autocorrelation has Bessel-like shape (see Fig. 2d) and its Fourier Transform is ring shaped with a strong peak in $k_x, k_y = 0$ as shown in Fig. 2e. Therefore no signature of hyper uniformity is present in these data. We will show below how to retrieve a hyperuniform pattern starting from the experimental data of Fig. 2c.

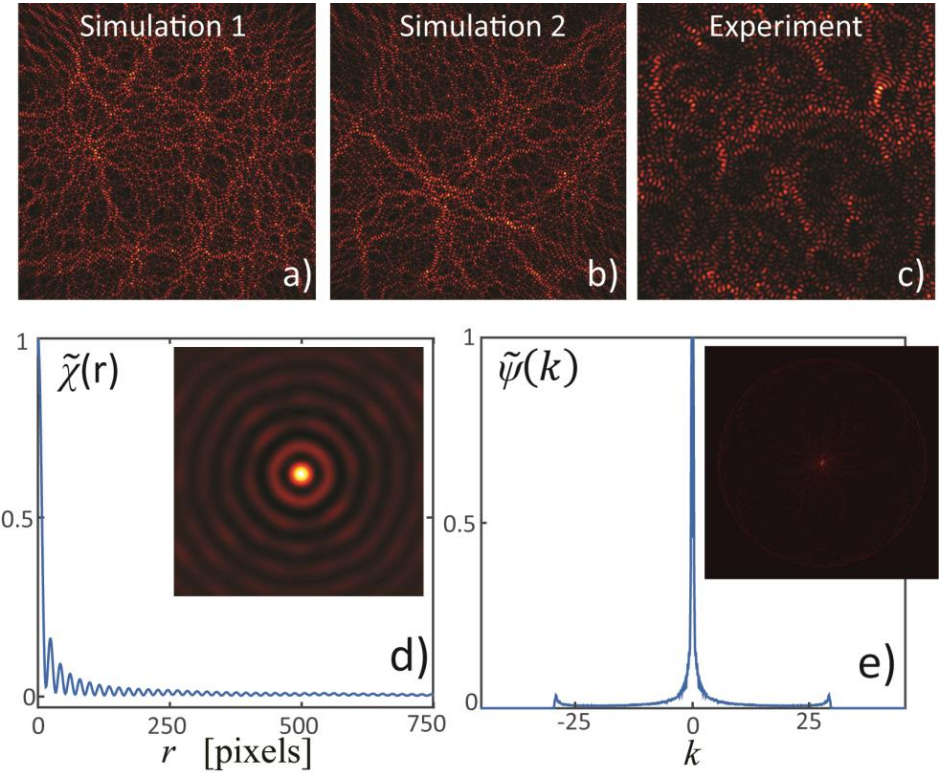

Fig. 2. ASPs are given by the superposition of Bessel beams. In panel a) we show the pattern from random superposition of Bessel beams, in b) the phasors are distributed according to the given hyperuniform map ($\chi = 0.4$). In panel c) we report the ASP from the experiment. In all the cases the curves of $\tilde{\chi}(r)$ and $\tilde{\psi}(k)$ are equivalent. The $\tilde{\chi}(r)$ is Bessel shaped as shown in d), the width of the first peak gives the typical speckle grain size, while the distance between the peak provides the average distance $\tilde{d}$ within the grains. In panel e) the $\tilde{\psi}(k) \neq 0$ in the annular region defined by $\tilde{k} = \frac{2\pi}{\tilde{d}}$ and around $k_x, k_y = 0$ where is strongly peaked. Outside the annular region $\tilde{\psi}(k) = 0$.

## 2.2 Analysis of the synthetic speckle grains distribution: a comparison between random speckle pattern and amorphous speckle pattern

Here we describe the process of retrieving a hyperuniform pattern starting from the position of the intensity maxima from an amorphous pattern as the one shown in Fig. 2c. Using a custom data analysis software we are able to find the local maxima of our speckle pattern and to obtain the position and the number of the speckle grains in each pattern. Similar analysis has been already adopted in previous works [49].

In such a way we are able to obtain a points' map that provides the coordinate/position of the grains from the speckle patterns and then analyse the properties of their distribution.

We extract the speckle grains coordinates from the two different conditions:
1) the standard speckle pattern (random-Gauss dephasor Fig. 3a)
2) the ASP (Bessel dephasor Fig. 3b).

The software pinpoints each intensity peak (speckle grain) with yellow crosses as shown in the panels a. and b. and extrapolate their coordinates. Those coordinates serve as maps to form the two patterns in panels c. and d. of Fig. 3 corresponding to intensity peak distribution of standard speckle patterns and ASPs respectively and the correspondent structure factors $S(k)$ are studied. While for the case related to conventional speckle pattern the $S(k) \neq 0$ for $k \to 0$ as shown in Fig. 3e, in the case of ASP shown in Fig. 3f the curve $S(k)$ tends to 0 when $k \to 0$, proving that the ASPs exhibit hyperuniform information intrinsically contained in the speckle grain distribution.

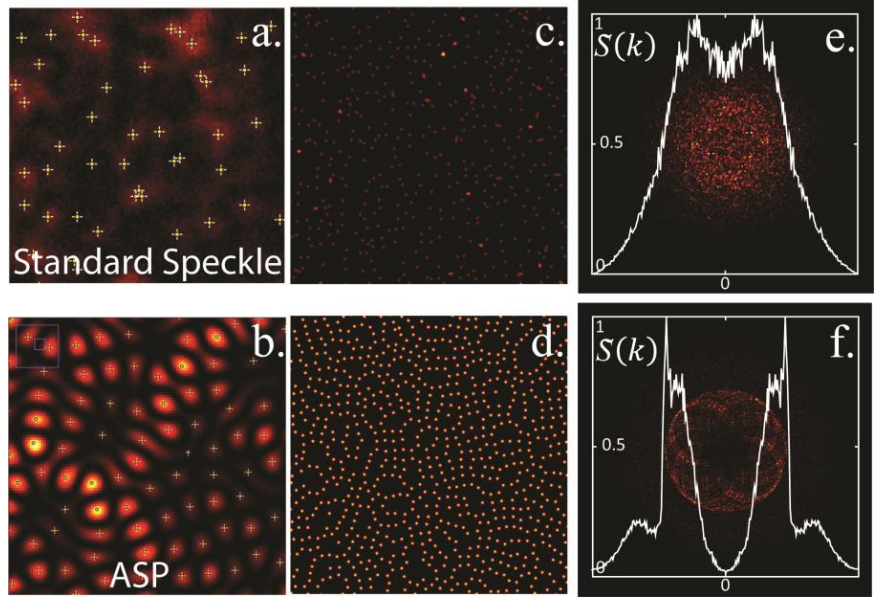

Fig. 3. Study of the distribution of the speckle pattern grains. We extrapolate the position of the speckle grains from a conventional speckle pattern (Gauss modes) in panel a. and a synthetic ASP (superposition of Bessel modes) in panel b.. Yellow crosses in patterns a. and b. define intensity peaks (speckle grains) positions. Peaks position are used as maps to generate point distributions in panel c. and d.. Their $S(k)$ are shown in panel e. for standard speckles and f. for ASPs. we notice that $S(k) \to 0$ when $k \to 0$; the speckle grains (intensity peaks) distribution of ASPs respect the condition for hyperuniformity.

In other words, the sufficient condition to obtain intensity maxima with a hyperuniform distribution is to use Bessel beams as sources, therefore they will be the first candidates to produce *hyperuniform continuous fields*. Gaussian sources provided the same result but only by binding their locations accordingly with a hyperuniform map.

## 3. The generation of the hyperuniform continuous fields from experimental speckle patterns

Here we demonstrate that it is possible to obtain the hyperuniform continuous field starting from experimental amorphous speckle and with a non-perturbative data treatment, in order to compensate for the intensity fluctuation encountered in the patterns.
We define $A$ as the two-dimensional matrix with $n_1 \times n_2$ elements representing the ASP (Fig. 4a) obtained with our experimental apparatus.
We generate a two-dimensional filter h of the Gaussian type: a rotationally symmetric Gaussian low-pass filter of size $h_{size}$ with positive standard deviation ($\sigma$). We used $h_{size}$ [$n_1$ $n_2$] with same size as the speckle pattern matrix and $\sigma = 20$ pixels.
We created the Gaussian filters using the following equations:
$$h_g(n_1, n_2) = e^{-(n_1^2 + n_2^2)/2\sigma^2} \qquad (7)$$

$$h(n_1, n_2) = \frac{h_g(n_1, n_2)}{\Sigma_{n_1} \Sigma_{n_2} h_g} \tag{8}$$

We filter the matrix $A$ (the original speckle pattern) adopting the filter h in such a way to generate the envelope $B$ as follows:

$$A^* = FT\{A\} \cdot h \qquad B = FT^{-1}\{A^*\} \tag{9}$$

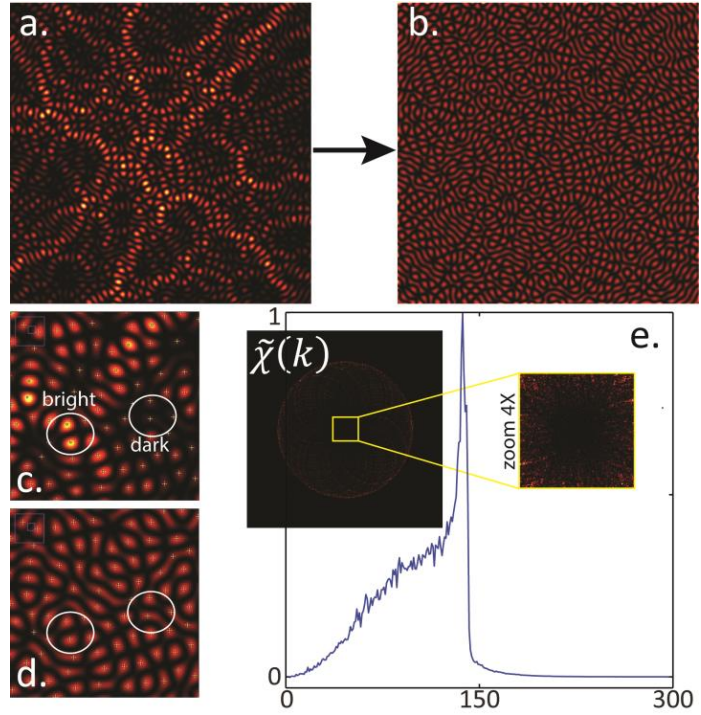

Figure 4: In panel a. the typical ASP, in b. the same pattern is corrected with a Gaussian filter. In c. and d. we show respectively 5X magnification from the patterns in a. and b., after correction the pattern loses the intensity fluctuations, e.g. the bright and dark regions in the white circles. The small white crosses in panels c. and d. indicate the speckle grains positions, it is visible that the Gauss-filter does not modify the grains positions but correct for their intensity only. The $\tilde{\chi}(k)$ from b. is shown with the inset in e. and the blue curve is its intensity profile. We have that $\tilde{\chi}(k) \to 0$ when $k \to 0$.

We normalize the original speckle pattern $A$ (Fig. 4a) dividing it with its envelope $B$ and we obtain the intensity corrected speckle pattern $C$ (Fig. 4b):

$$C(l, m) = A_{l,m}/B_{l,m} \qquad \forall l \in [1, n_1], \forall m \in [1, n_2] \tag{10}$$

Comparing $A$ (panel a. in Fig. 4) and $C$ (panel b. in Fig. 4) we notice that they differ in their intensity distribution only as shown in the insets c and d of Fig. 4, zoom in from panels a. and b. In $C$ we observe intensity fluctuations (extraordinary peaks) as shown with the white circles in the Fig. 4c and 4d. We retrieved the speckle grains positions in $A$ and $C$ and we calculated their cross correlation between the two maps obtaining the correlation coefficient 0.88 indicating that speckle grains position are not modified by the filtering process as expected.

In practice, adopting the filtering process we eliminated the intensity fluctuations from $A$ and we obtain the "homogeneous" pattern shown in panel b. of Fig. 4, the intensity corrected amorphous speckle pattern. If we analyse the statistics of the corrected pattern $C$ we have that $\tilde{\chi}(k) \to 0$ when $k \to 0$ as shown in section e. of Fig. 4, i.e. it is hyperuniform.

## 3.1 Local Field Variance

In order to further confirm the hyperuniform distribution of ASPs we calculate the local scalar field fluctuations $R^2\sigma_F^2(R)$ of speckles within a circular observation window of radius R [39, 40].
The speckle patterns analysed are rasterized into a window with linear size W, where W = 400pixels according to the original image linear size.

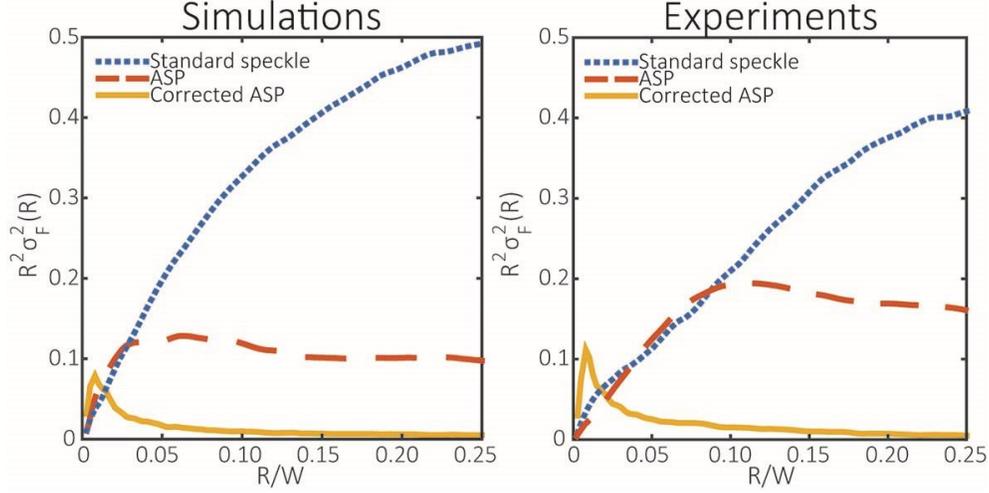

Figure 5: On the left panel the comparison of the local scalar field fluctuation versus the windows radius from different speckle patterns: dotted blues curve from synthetic standard speckle pattern, dashed red curve from the synthetic ASP and the intensity corrected ASP (the one in Fig.4b). On the right panel the same comparison is presented for the experimental data.

Thereafter, we define a sub-window with variable radius R ranging from 2 pixels up to the maximum window radius $R_{max}$, where $R_{max} = \frac{W}{4} = 100$pixels. We count the total intensity found within the region of interest (the sub-window) and we store this data while scanning with the sub-window the whole mask. The scan is performed in step of 10 pixels in both directions x and y in the mask. Boundary conditions are taken into account as the mask would have been replicated in each direction, in this way if the sub-window exits the mask on one side it emerges from the opposite. This allows to estimate the average intensity (the volume) residing in each sub-window size (from 2 pixels to $R_{max}$) and calculate the corresponding volume fraction variance. The results for each of the speckle pattern analysed is shown in the plot in Fig. 5.
In the left panel of Fig. 5 we compare the volume fraction variance calculated for synthetic ASP (red dashed curve), the synthetic random speckle pattern (blue dotted curve) and the intensity corrected ASP (yellow curve). We observe that while for the random distribution $R^2\sigma_F^2(R)$ tends to diverge with R, it tends to decay for the ASP. On the contrary, in case of the intensity corrected ASP $R^2\sigma_F^2(R)$ decays to 0 with R, as required for hyperuniform distributions.
Indeed, the amorphous speckle pattern is characterized by the hyperuniform distribution of the speckle grains, in practice the number variance $\sigma_N^2(R)$ of the ASP's grains within an observation window of size R grows more slowly than the sub-windows volume in the large-R limit, i.e., slower than $R^d$ (in our case d = 2). On the contrary, for the standard speckle pattern grains distribution $\sigma_N^2(R)$ grows quadratic with R (i.e., non-hyperuniform).

The same comparison is shown in the right panel of Fig. 5 for the experimental patterns, which appear in agreement with the numerical simulation.

## 4. Summary and Outlook

By filtering with an annular aperture the spatial frequencies of a random distribution (in our case a standard speckle pattern) we obtain an amorphous pattern. The ASP emerges from the superposition of Bessel beams and always exhibits hyperuniformity in its speckle grain distribution. We demonstrate that the maxima of the ASP distribution are arranged in hyperuniform fashion. Moreover starting from the ASP it is possible to produce (after proper intensity filtering) an intensity distribution fulfilling requirements for hyperuniformity and consisting of a random superposition of Bessel beams.

We have proved that our method discloses the possibility to generate continuous hyperuniform fields on optical bench, light distributions that can be fully tailored with commercial phase modulators. It follows that hyperuniform light fields can be used to form engineered photonic architectures capable of strong light confinements [24, 30].

Our method has potential application for multi photon fabrication methods in which polymerization only happens above a threshold [50, 51]: in such case high intensity speckle grains maxima can be employed to produce a hyperuniform defects distribution. Ultimately, we believe that the work presented will inspire new forms of complex light patterns of interest for several applications ranging from ultracold atoms to optical imaging.

## 5. Methods and Validations

### 5.1 Description of the experimental systems

In our experiments speckle patterns are produced with the setup depicted in Fig.6a. The beam emitted at 633nm wavelength by a He-Ne laser is reduced to 1mm in waist by the iris (F) and impinges on to a ground diffuser (D). Trespassing D, the light experiences multiple scattering and results completely scrambled in phases and directions. A 4f system composed of lenses L1 (focal length f1=25,4mm) and L2 (focal length f2=250mm) magnifies 10X the plane at 1mm from the back surface of D and projects it onto the CCD camera. It results the speckle pattern $C_{experiment}$ reported in the right panel of Fig. 7b.

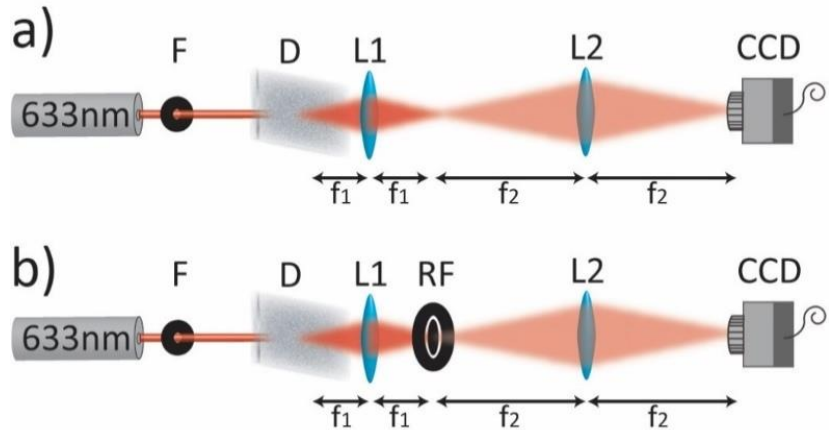

Fig. 6. In section a) the scheme of the experimental setup is depicted. A coherent laser beam impinges onto a scattering diffuser (D). The transmitted light at 1mm at the back of D is projected with a 4f system composed of lenses L1 and L2 and is magnified 10X on the CCD camera plane. In section b) we apply a spatial frequency selection of the forward scattered light introducing an annular aperture (ring filter RF) in the Fourier plane of the 4f system. In this last case the ASP is generated.

In order to generate ASPs, we apply a spatial frequency selection to the light transmitted through D, a filtering operation performed by an annular aperture. As depicted in Fig. 6b, we center on the Fourier plane of the 4f system (25,4mm at the back of L1) a commercial ring filter (RF). The RF is the R1CA2000 annular aperture from Thorlabs with 1700μm obstruction diameter and 2000μm pinhole diameter. Exploiting a ring shaped filter we generate a superposition of plane waves with fixed k and random phase, which is known to give rise to an ASP as the one shown in Fig. 2c.

## 5.2 The numerical method and its analysis

The speckle phenomenon results from the superposition of a multitude of randomly phased elementary components (dephasor) and can be simulated implementing a simple numerical model that sums the dephasors on a hypothetical observation plane [48]. Within this introductory section we validate our numerical model, and we aim to compare the speckle pattern resulting from our simulations with those from the experiments considering their statistics [48]. First, we simulate the light speckles formation, the intensity pattern on the camera plane (the observation plane), may be described as a random superposition of Gaussian components.

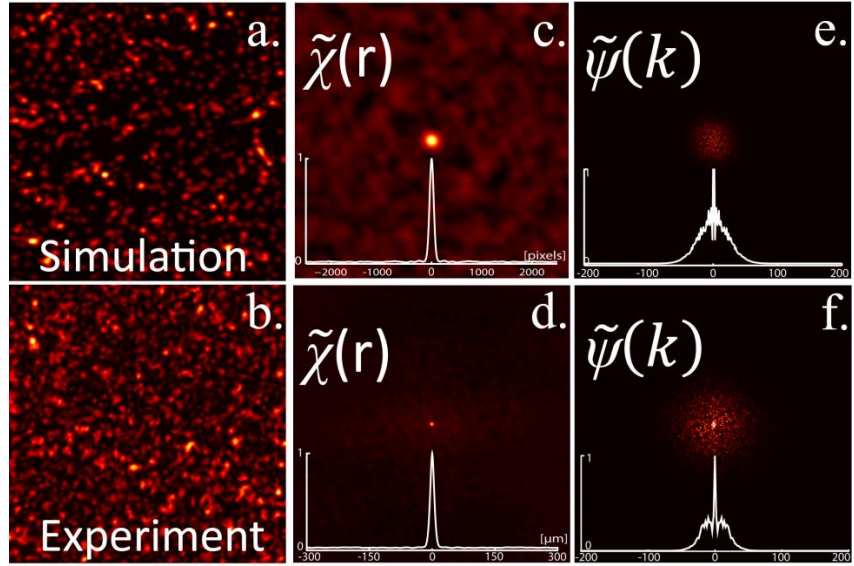

Fig. 7. Standard speckle patterns analysis. We show the speckle pattern from the numerical simulation in panel a and experimental speckle patter panel b. In panels c and d we report their correspondent auto-correlation function $\tilde{\chi}(r)$ and in e and f their spectral density $\tilde{\psi}(k)$. Comparing the two cases we prove that our simulations respect the statistic from the real experiment.

Following this assumption, in our simulation we impose a superposition of Gaussian modes. Given a [X,Y]=5000x5000 pixels grid, a single mode can be described as follow:

$$f_i = e^{-\frac{(x-x_i)^2 + (y-y_i)^2}{2\sigma^2}} \cdot e^{i\cdot\varphi_i} \qquad (11)$$

where the point $(x_i, y_i)$ determines the position of the i-th mode on the grid and $\sigma$ defines the width of the Gaussian mode. If FWHM is the Full Width Half Maximum of the mode, then $\text{FWHM} = 2\sqrt{2\ln 2}\,\sigma$. To each of the mode we associate a random term of phase $\varphi_i$ with $\varphi_i \in [0, 2\pi]$ and random positions $R_i = (x_i, y_i)$. Thereafter, to calculate the final speckle pattern we superimpose the $i = N$ modes on the grid with fixed $\sigma$ using the following formula:

$$C_{simulation} = \left(\sum_{i=1}^{N} f_i\right)^2. \qquad (12)$$

The derived synthetic pattern $C_{simulation}$, shown in Fig. 7a, can be directly compared with the speckle pattern $C_{experiment}$ (Fig. 7b) obtained from the experiment.

We calculate their auto-correlation function, $\chi(r)$, as an inverse Fourier transform of the speckle (C) energy-spectrum:

$$\chi(r) = \text{FT}^{-1}\{|\text{FT}\{C - \mu_C\}|^2\} \qquad (13)$$

and where $\mu_C$ is the mean value of C. This procedure is commonly used for phase retrieval imaging together with the speckle envelope correction [44]. In addition, we take the volume of the system to be unity, therefore we always consider the normalized quantity $\tilde{\chi}(r)$ defined as $\tilde{\chi}(r) = \chi(r)/\chi(0)$, where $\chi(0)$ is the auto-correlation maxima in $r = 0$. The Fourier Transform (FT) of $\tilde{\chi}(r)$ corresponds to the spectral density $\tilde{\psi}(k)$ that is calculated as follows:

$$\tilde{\psi}(k) = \text{FT}\{\tilde{\chi}(r)\} \qquad (14)$$

The two patterns are equivalent as proved with the comparison of their statistical analysis shown in Fig. 7, where their auto-correlation function $\tilde{\chi}(r)$ (panels c and d) and its Fourier Transform, the spectral density $\tilde{\psi}(k)$ (panels e and f), are considered.

For both the patterns the $\tilde{\chi}(r)$ shows a single central peak, its width is equivalent to the typical speckle grain size of the pattern. On the contrary, $\tilde{\psi}(k)$ describes the spatial frequencies distribution which the speckle pattern is composed of. In general, for standard speckle patterns the spatial frequencies are distributed around $k_x, k_y = 0$ (the peak) and decay for larger k as shown in Fig. 7e and 7f.

**Acknowledgment**

We wish to acknowledge Prof. Salvatore Torquato and His graduate students Duyu Chen, Jaeuk Kim and Zheng Ma from Princeton University (New Jersey, USA) for providing us with hyeruniform matrix from Ref. [34] and their valuable feedbacks during the development of this work. This work was supported by the EU FP7 Marie Curie ITN "OILTEBIA" PITN-GA-2012-317526 and the H2020 "Laserlab Europe" (EC-GA 654148).